\documentclass[twocolumn,prb,aps]{revtex4}
\usepackage{graphicx}
\usepackage{graphics}
\usepackage{epsfig}

\begin{document}

\title{Electron collisions with cyanoacetylene HC$_3$N: vibrational excitation and dissociative electron attachment}
\author{M. Rankovi\'{c}$^1$, P. Nag$^1$, M. Zawadzki$^{1,2}$, L. Ballauf$^3$, J. \v{Z}abka$^1$, M. Pol\'{a}\v{s}ek$^1$, J. Ko\v{c}i\v{s}ek$^1$, J. Fedor$^{1}$}	
\email{juraj.fedor@jh-inst.cas.cz}
\affiliation{$^1$J. Heyrovsk\'{y} Institute of Physical Chemistry, Czech Academy of Sciences, Dolej\v{s}kova 3, 18223 Prague, Czech Republic}
\affiliation{$^2$Department of Atomic, Molecular and Optical Physics, Faculty of Applied Physics and Mathematics, Gda\'{n}sk University of Technology, ul. G. Narutowicza 11/12, 80-233 Gda\'{n}sk, Poland}
\affiliation{$^3$ Institute of Ion Physics and Applied Physics, University of Innsbruck, Technikerstrasse 25, A-6020 Innsbruck, Austria}

%\date{}

\begin{abstract}
We experimentally probe electron collisions with HC$_3$N in the energy range from 0 to 10 eV with the focus on vibrational excitation and dissociative electron attachment. The vibrational excitation cross sections show a number of resonances which are mode specific: the two dominant $\pi^*$ resonances are visible in the excitation of all the vibrational modes, however, broad $\sigma^*$ resonances are visible only in certain bond-stretching vibrational modes. The lower $\pi^*$ resonance shows a pronounced boomerang structure. Since it overlaps with the threshold peak originating from a long-range electron-molecule interaction, the interference pattern is rather unusual. Somewhat surprisingly, the boomerang structure is visible also in the elastic scattering cross section. The dissociative electron attachment cross sections agree qualitatively with the data of Gilmore and Field [J. Phys. B 48 (2015) 035201], however, approximately a factor of two difference is found in the absolute values.
\end{abstract}

\maketitle

\section{Introduction}
Cyanoacetylene, HC$_3$N, %has been detected spectroscopically in a number of extraterrestrial environments. 
has been attracting attention due to its abundance in a number of extraterrestrial environments.
Among these are interstellar clouds,~\cite{turner71} circumstellar envelopes,~\cite{bieging93} comets~\cite{irvine81} and atmosphere of Saturn's moon Titan.~\cite{kunde81, coates07}
%It is believed that the chemical changes in some of these environments are to a large degree driven by the impact of low-energy electrons. These electrons  %It's spectroscopic detection is facilitated by 
 %it has been detected by spectroscopic methods in  the coma of comet Hale–Bopp and in the atmosphere of Saturn's moon Titan, where it sometimes forms expansive fog-like clouds. 
The particular interest in the electron collisions with this molecule stems primarily from two sources. The first is the presence of the carbon-chain molecular anions such as C$_8$H$^-$, C$_6$H$^-$, C$_4$H$^-$ and C$_3$N$^-$  in the interstellar medium.~\cite{brunker07, cernicharo07, thaddeus08} The second is the 2007 observation of the Cassini mission,~\cite{coates07} that the upper atmosphere of Titan contains anions with mass/charge ratio of up to $\approx$ 10000. Extensive investigations have shown, that depending on the altitude, the dominant anion species in Titan's atmosphere are either CN$^-$ and C$_3$N$^-$, or C$_n$H$^-$, with $n = 2, 4, 6$.~\cite{vuitton09}

The dissociative electron attachment (DEA) to neutral polyynes (HC$_n$H or HC$_n$N) as a possible dominant source of these anions has been ruled out early. The DEA studies to C$_2$H$_2$,~\cite{may_acet09} C$_4$H$_2$,~\cite{may_diac08} and HC$_3$N~\cite{graupner06} have shown that while the cross sections are considerably high, the fragmentation channels are endothermic. The energetic thresholds for the production of fragment anions lie in all these cases above 1~eV, and are thus inaccessible for thermal electrons. Nonetheless, a formation of transient anions - resonances - leads not only to DEA but due to competing electron autodetachment channel also to vibrational excitation of the molecules. This influences both the vibrational energy distribution of the gas and the electron energy distribution function in the above-mentioned astrochemical environments. 

The only electron collision experiments with HC$_3$N to our knowledge are the early positive and negative ionization studies of Dibeler~\cite{dibeler60} and Harland~\cite{harland86} and the DEA experiments in the group of T. Field, QU Belfast.~\cite{graupner06, gilmore15} The latter group has initially reported a yield of individual fragment ions~\cite{graupner06} and later recalibrated these yields using signal from background water vapor to determine the absolute partial cross section values.~\cite{gilmore15} Theoretically, the resonances in cyanoacetylene were explored by Sommerfeld and Knecht~\cite{sommerfeld05} with the complex absorbing potential approach, by Sebastianelli and Gianturco~\cite{sebastianelli12} with the single-center expansion scattering calculations and by Kaur et al.~\cite{kaur16} by R-matrix theory. Orel and Chourou~\cite{orel_hc3n11} performed multidimensional nuclear dynamics calculations on the resonant states of HC$_3$N. 

In the present paper we probe the resonant states in cyanoacetylene by the means of electron energy loss spectroscopy. We report the absolute differential elastic and vibrationally inelastic cross sections at 135$^\circ$ scattering angle. These measurements bring detailed information about the resonant electronic states and the dynamics of the nuclear motion on their potential energy surfaces. The observed selectivity in the excitation of certain vibrational modes facilitates the assignment of the involved resonances. We also report direct absolute measurement of the DEA cross section. 

\begin{table*}
\begin {center}
	\caption{Unoccupied molecular orbitals of neutral HC$_3$N and corresponding resonance energies formed by capture of an electron into the orbital (in eV).}
	\label{tab:orbitals}
\begin{ruledtabular}
\begin{tabular}{lcllll}
Symmetry & MO isosurface & Present scaling  & CAP~\cite{sommerfeld05}  &  Scattering calc.~\cite{sebastianelli12} & R-matrix~\cite{kaur16} \\
\hline
%\vspace{0.5cm}
$\pi_1^*$ & \raisebox{-0.5\totalheight}{\includegraphics[width=4cm]{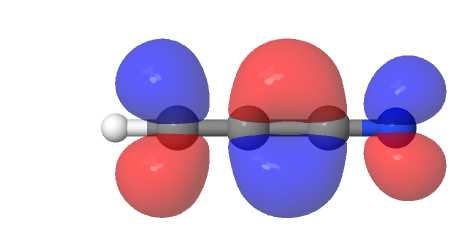}}& 0.48 & 0.7 & 1.94 & 1.51 \\ 
$\sigma_1^*$ & \raisebox{-0.5\totalheight}{\includegraphics[width=4cm]{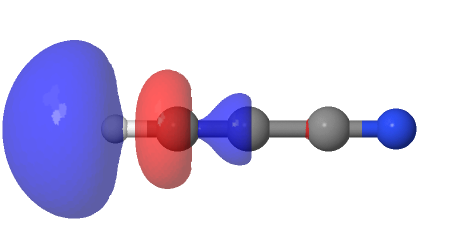}} & 3.09& & &  \\
$\pi_2^*$ & \raisebox{-0.5\totalheight}{\includegraphics[width=4cm]{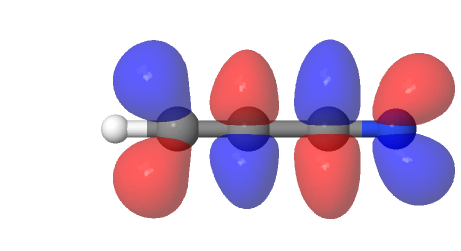}} & 5.50& 6.2 & 8.19 & \\ 
$\sigma_2^*$ & \raisebox{-0.5\totalheight}{\includegraphics[width=4cm]{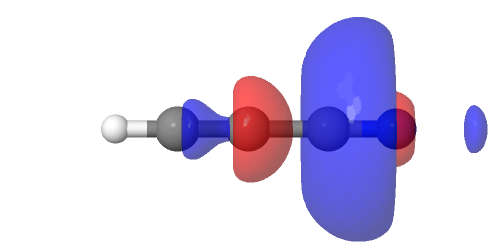}} & 5.39 & & 9.24 & 8.0 \\
 \end{tabular}
 \end{ruledtabular}
\end {center}
\end{table*}

\section{Experiment}
Three electron-collisions setups were used for the present experiments, recently transferred to Prague from the University of Fribourg.

The electron scattering experiments were performed on the electrostatic spectrometer with hemispherical electron monochromator and analyser.~\cite{allan_ELS92, allan_ELS05} The electrons scattered on the effusive beam of the pure sample gas were analysed at the fixed scattering angle of 135$^\circ$. The energy of the incident beam was calibrated on
the 19.365 eV 2$^2$S resonance in helium. Electron-energy resolution was 17 meV. The absolute elastic scattering cross section was calibrated against the one of helium using a relative flow method. %The setup operated in three different modes: either 
The detailed error budget of the cross section calibration has been presented in Ref.~\onlinecite{allan_thf07}. The uncertainly of the elastic cross section is $\pm$15\%. The vibrationally inelastic cross sections are normalized with respect to the elastic peak. Since the individual vibrational modes are not fully resolved, the individual vibrational excitation cross sections are much less precise and should be considered as indicative values, which describe the intensity of the inelastic signal at a given energy loss

The absolute dissociative electron attachment cross sections were measured on the absolute DEA spectrometer with time-of-flight mass analyzer.~\cite{may_acet09, may_diac08} A pulsed magnetically collimated electron beam, produced in a trochoidal electron monochromator, crosses collision cell filled with a stagnant gas and the anions produced are extracted towards short (15 cm) time-of-flight mass analyzer placed perpendicularly to the electron beam. For the cross section calibration, we have used the 4.4~eV band in the  O$^-$ production from CO$_2$ with the energy-integrated cross section of 13.3 eV pm$^2$. The same band is used for the electron energy scale calibration and for the determination of the electron beam resolution which was $\approx$ 250~meV. The uncertainty of the absolute DEA calibration is $\pm$ 20\% which includes both the systematic and statistical errors.

The shape of the DEA bands was additionally measured on the DEA spectrometer with a trochoidal monochromator and quadrupole mass filter~\cite{stepanovic99, langer18, zawadzki_pyruvic18}. Here, a continuous electron beam crosses the effusive molecular beam and the yield of a certain anion mass chosen by the quadrupole is monitored. Due to absence of pulsing, this spectrometer has a better electron energy resolution of approximately 100~meV. The final DEA cross sections are thus obtained by scaling the high-resolution DEA yields from the quadrupole setup to the absolute values from the time-of-flight setup using the invariance of the energy-integrated cross sections.~\cite{janeckova_formic13, graupner_ccl2f210}

The HC$_3$N sample was synthesized by the dehydration of the propiolamide, prepared by the reaction of methylpropiolate and amonia, the method introduced by Miller and Lemmon~\cite{miller67}. During the measurements, the sample (confined in a lecture bottle) was kept at the temperature of 7~$^\circ$ C.  

\section{Results and discussion}
\subsection{Electronic structure and resonances}
All three scattering  processes probed in this work are strongly influenced by the formation of resonances - temporary anion states - in the electron molecule collision. We thus first review the available information on these states, which will facilitate the interpretation of the results and further discussion. 

Since the resonant states are embedded in continuum, their proper characterization requires advanced scattering calculations or modifications of the traditional quantum chemistry approaches. However, a useful insight can be gained from the basic electronic structure of the target molecule and a use of the scaling formulas. In a simplified picture a shape resonance can be imagined as trapping of the incident electron in an unoccupied molecular orbital of the target molecule. Cyanoacetylene is a linear polyyne with two triple bonds. The lowest four unoccupied orbitals, shown in the table~\ref{tab:orbitals}, have antibonding character along some, or all bonds. For the purpose of this paper we denote them $\pi_1^*, \pi_2^*$ and $\sigma_1^*$, $\sigma_2^*$.  Chen and Gallup~\cite{chen90} developed an empirical scaling based on the Koopmans' theorem, relating the orbital energies ($E_{MO}$) and the corresponding resonance energies ($E_{res} = (E_{MO}$ - 2.33~eV) / 1.31). Values obtained using this formula are listed in table~\ref{tab:orbitals} in the ``present scaling'' column. It should be noted that the sensitivity of such estimate of $E_{res}$ to the choice of basis set and the scaling formula have been explored by Field and co-workers.~\cite{graupner06, millar17}
%The basic idea of the scaling is to calculate the orbital energies with a very small basis set (6-31G$^*$) that artificially confines the 
%Figure~\ref{fig:orbitals} shows the four lowest unoccupied molecular orbitals of HC$_3$N, the corresponding $E_{MO}$ and $E_{res}$. There are two $\pi^*$ resonances predicted at 0.48 and 5.5~eV and two $\sigma^*$ resonances at 3.09 and 5.35~eV. 

The resulting resonant energies can be considered only as indications, however, as can be seen in table~\ref{tab:orbitals} they agree surprisingly well with the advanced theoretical approaches. The complex absorbing potential (CAP) method of Sommerfeld and Knecht predicted the $\pi_1^*$ resonance at 0.7~eV (width 0.15~eV)
and the $\pi_2^*$ resonance at 6.2~eV (width 1.1~eV). The scattering calculations of Sebastianelli and Gianturco localized the $\pi^*$ resonances at somewhat higher energies of 1.94~eV (width 0.15~eV) and 8.19~eV (width 0.76~eV) and the $\sigma_2^*$ resonance at 9.23~eV (width 1.16~eV). The R-matrix calculations of Kaur et al. identified the $\pi_1^*$ at 1.51 eV and $\sigma_2^*$ at 8~eV. 
An alternative scaling formula developed recently by Field and co-workers especially for $\pi^*$ states in conjugated systems, predicts the two resonances at 0.5 and 5.1~eV. 

Two notes should be added at this point. First, the figures in table~\ref{tab:orbitals} are the isosurfaces of the  molecular orbitals, i.e. unoccupied one-electron states. %, so they have in principle nothing to do with the scattering wavefunction. 
Sebastianelli and Gianturco~\cite{sebastianelli12} provided the graphical representations of the true one-electron scattering wave functions and they are very similar (basically indistinguishable by eye) to the present isosurfaces. This adds the credit to the simplified picture of the temporary orbital occupation by the incoming electron. The nodal planes and electron densities of the unoccupied orbitals will be  useful in interpreting the selectivity of vibrational excitation.  
The second note concerns the $\sigma^*$ states. The corresponding resonances are expected to be very broad: their coupling with the barrierless s-wave autodetachment channel leads to their extremely short lifetimes. The fixed-nuclei scattering calculations, which localize the resonances from the variation of the eigenphase sum, have thus often difficulties in finding such  broad resonances~\cite{AAMOP_chapter}: the eigenphase variation can be so weak that it is difficult to distinguish from the background scattering. This might be the case of the $\sigma_1^*$ resonance, lying between the two $\pi^*$ resonances, which was not reported in any of the scattering calculations. However, as will be shown below, this state is manifested in the vibrational excitation cross section of the C-H stretching mode. 
 
%They are thus difficult to be pinpointed in the scattering calculations, which is usually done from the shape of the eigenphase sums.  

\subsection{Elastic scattering}

\begin{figure}[tb]
\includegraphics[width = 7cm]{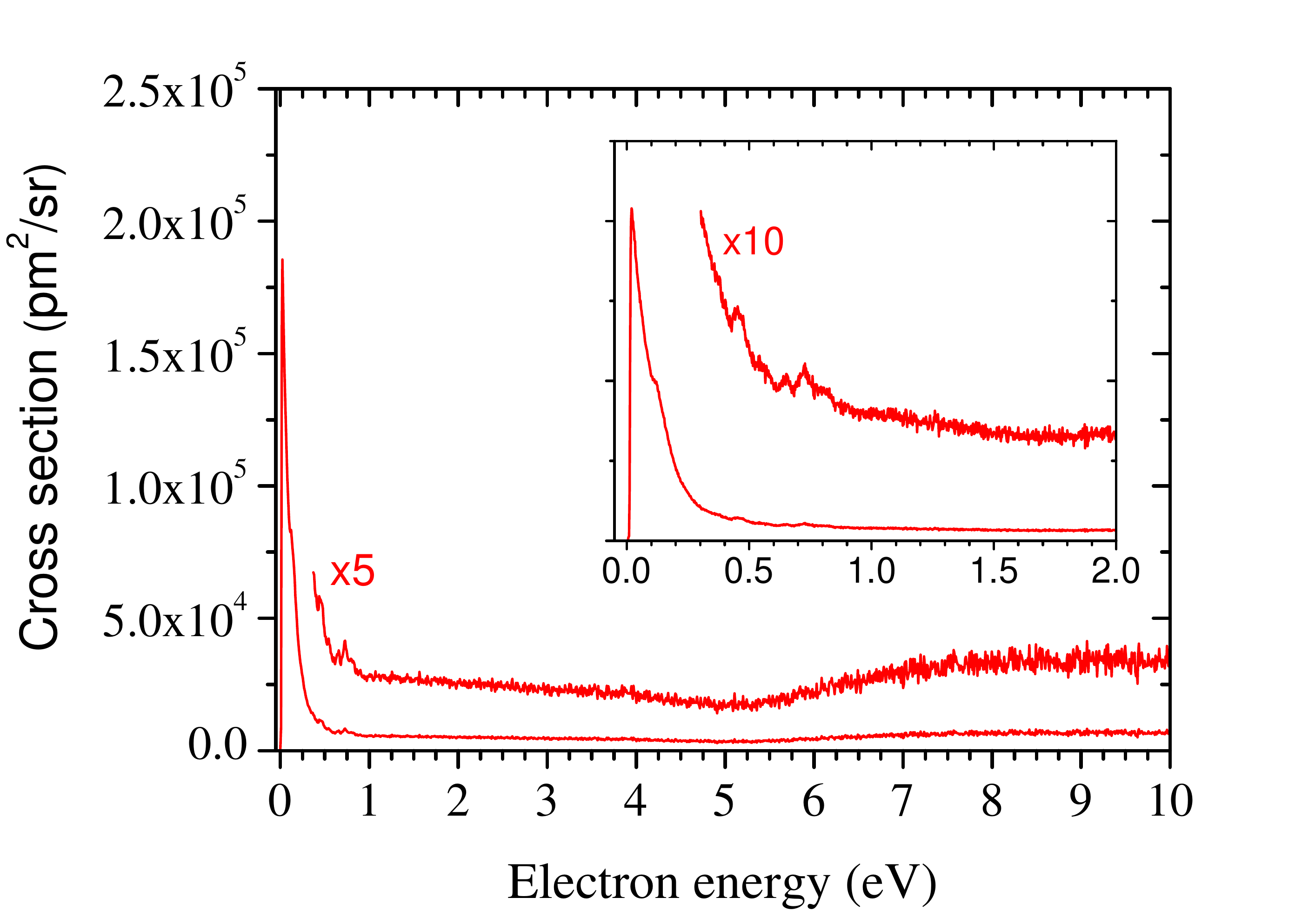}
\caption{Cross section for the elastic scattering on HC$_3$N at 135$^\circ$. The inset shows horizontally magnified electron energy scale.}
\label{fig:elas}
\end{figure}

Figure~\ref{fig:elas} shows the differential elastic electron scattering cross section at 135$^\circ$ scattering angle. The cross section sharply peaks towards 0~eV electron energy. This is caused by the dipole moment of HC$_3$N which is 3.72~Debye.~\cite{crc07} The elastic scattering cross sections in polar targets always reach high values, and in some cases even diverge, at very low energies.~\cite{fabrikant16} It should be noted that the true height of the low-energy spike is of course not accessible by a cross beam experiment such as the present one, since the monochromator and the analyzer can not reliably produce/analyze the electrons below some 30~meV kinetic energy.

Two interesting features can be observed in the cross section at higher energies. One is the shallow minimum around 5~eV. As shown in the next section, a broad $\pi_2^*$ resonance dominates this region and the minimum is an imprint of this resonance in the elastic cross section. Since its formation leads to increase in all vibrational excitation channels, the drop in the elastic channel is caused by the conservation of the probability flux.
The second interesting feature in the elastic cross section is the oscillatory structure between 0.4 and 0.8~eV. %\textbf{Before speculating: prove, that it cannot be due to overlap of the elastic peak and the 26~meV bending vibration}.
It is clearly connected with the threshold peaks and the $\pi_1^*$ resonance in the vibrational excitation cross sections in this energy range discussed below. %The fixed-nuclei scattering calculations of 
 Sebastianelli and Gianturco~\cite{sebastianelli12} and Kaur et al.~\cite{kaur16} have seen the influence of the resonances in the elastic scattering (in computed integral cross sections). However, since these were fixed-nuclei calculations, which do not reflect the probability flux towards the nuclear motion, the resonances were manifested as peaks in the cross sections, not as the dips observed here.  

\begin{figure}[tb]
\includegraphics[width = 7cm]{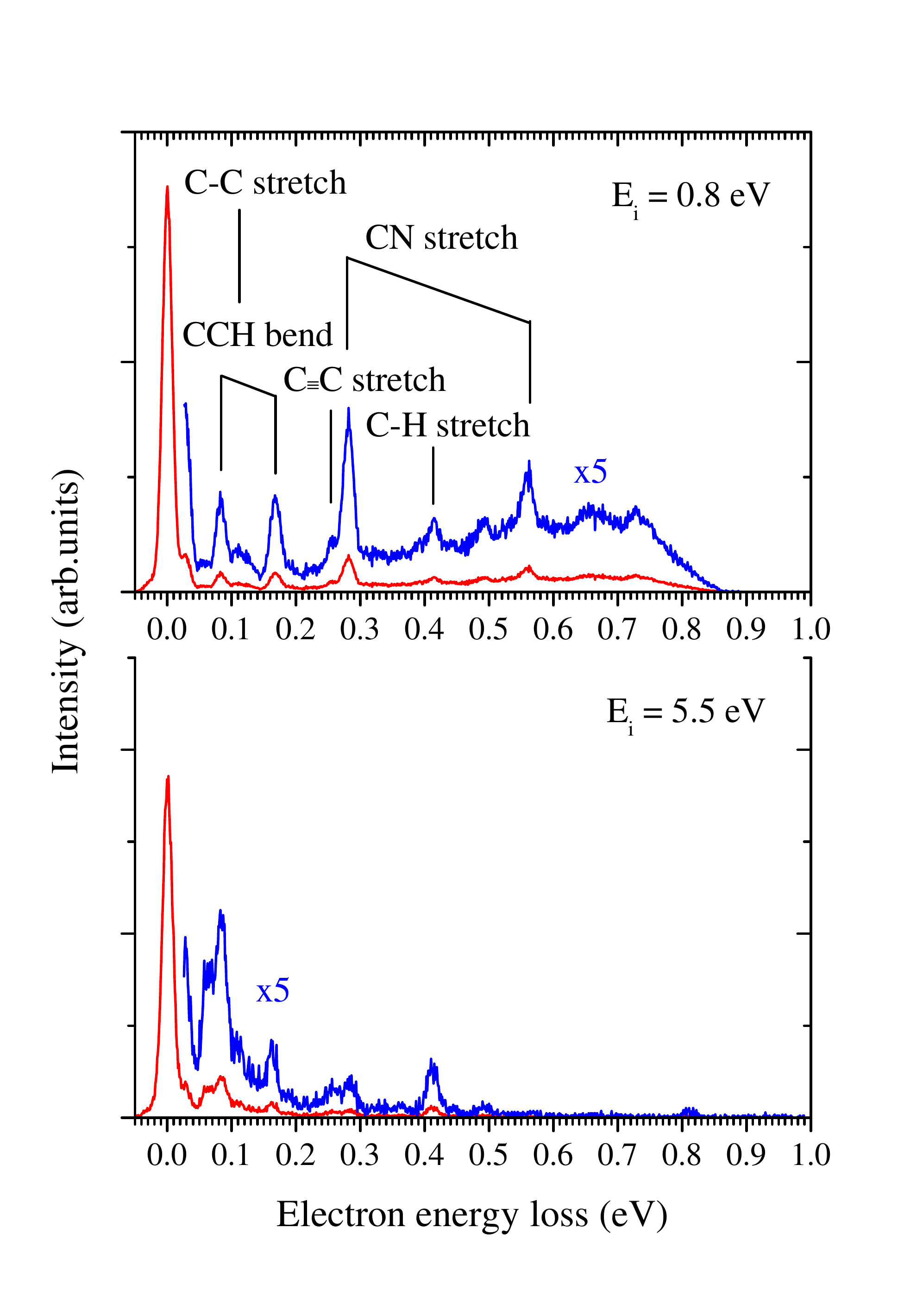}
\caption{Electron energy loss spectra of HC$_3$N at 135$^\circ$ recorded at incident energies of 0.8~eV (top panel) and 5.5~eV (bottom panel).}
\label{fig:eels}
\end{figure}

\subsection{Vibrational excitation}
\begin{table}
\begin {center}
\caption{Experimental vibrational frequencies of HC$_3$N from Ref.~\onlinecite{leach14}}
\label{tab:vibr}
\begin{tabular}{llll}
\hline
Type & Label & Energy (meV) \\
\hline
CCN bend & $\nu_7$ &  28 \\
CCC bend & $\nu_6$ &  62 \\ 
CCH bend  & $\nu_5$ & 82 \\
C$-$C stretch  & $\nu_4$ & 109 \\
C$\equiv$C stretch & $\nu_3$ & 257 \\
C$\equiv$N stretch & $\nu_2$  & 282 \\
C$-$H stretch  & $\nu_1$ & 412 \\
\hline

\end{tabular}
\end{center}
\end{table}

Figure~\ref{fig:eels} shows electron energy loss spectra recorded at two different electron incident energies. The energy loss spectra reflect, which vibrational modes are excited upon the electron impact and their relative population with respect to the elastically scattered electrons with zero energy loss. The spectroscopic experimental vibrational energies from Ref~\onlinecite{leach14}. are shown in table~\ref{tab:vibr}. 

All the three bending modes are excited to certain extent. The softest vibration, CCN bend (26~meV excitation energy) is visible as a shoulder of the elastic peak at both impact energies. The CCC bending vibration (62~meV) is not visible at 0.8~eV but present at 5.5~eV impact energy. The most prominent bending vibration is the CCH bend with the excitation energy of 82~meV, with at least one overtone excited at both incident energies (the possible $v=2$ overtone peak overlaps with the C$\equiv$C stretching mode). 
The excitation of the stretching modes also shows certain selectivity: at both incident energies, the C-C stretch is excited only weakly and the other vibrations have varying strength. At 0.8~eV, the C$\equiv$N stretch (282~meV) progression dominates the spectrum, while at 5.5~eV the C-H stretch becomes the dominant stretching mode. An interesting peak occurs at 492~meV (unassigned in the figure), which has to originate from a combination vibration of C-H stretch and CCH bend ($\nu_1 + \nu_5$). 

\begin{figure}[tb]
\includegraphics[width = 7cm]{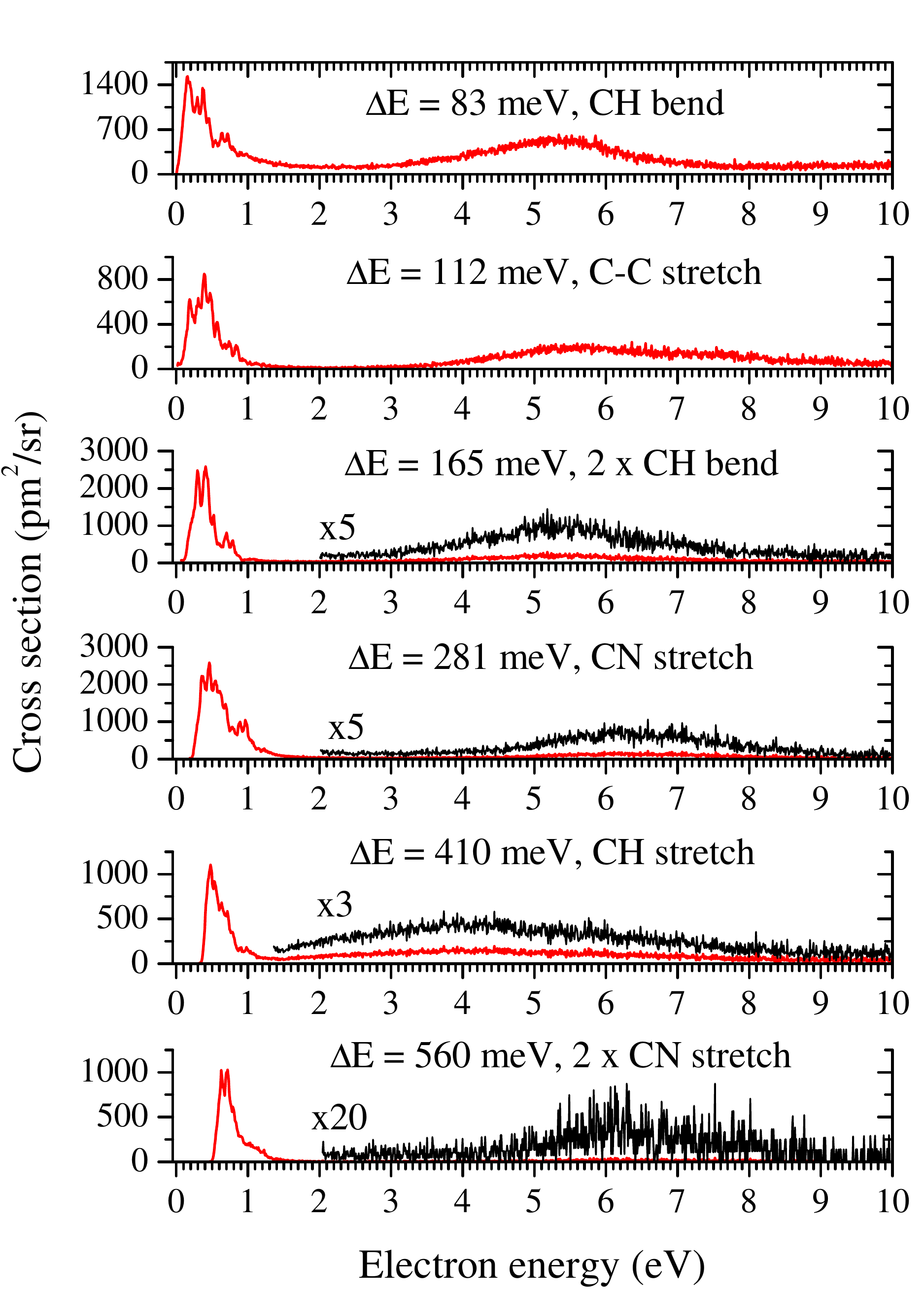}
\caption{The vibrational excitation cross sections for individual vibrations in HC$_3$N as functions of incident electron energy.}
\label{fig:eds_long}
\end{figure}

\begin{figure}[tb]
\includegraphics[width = 7cm]{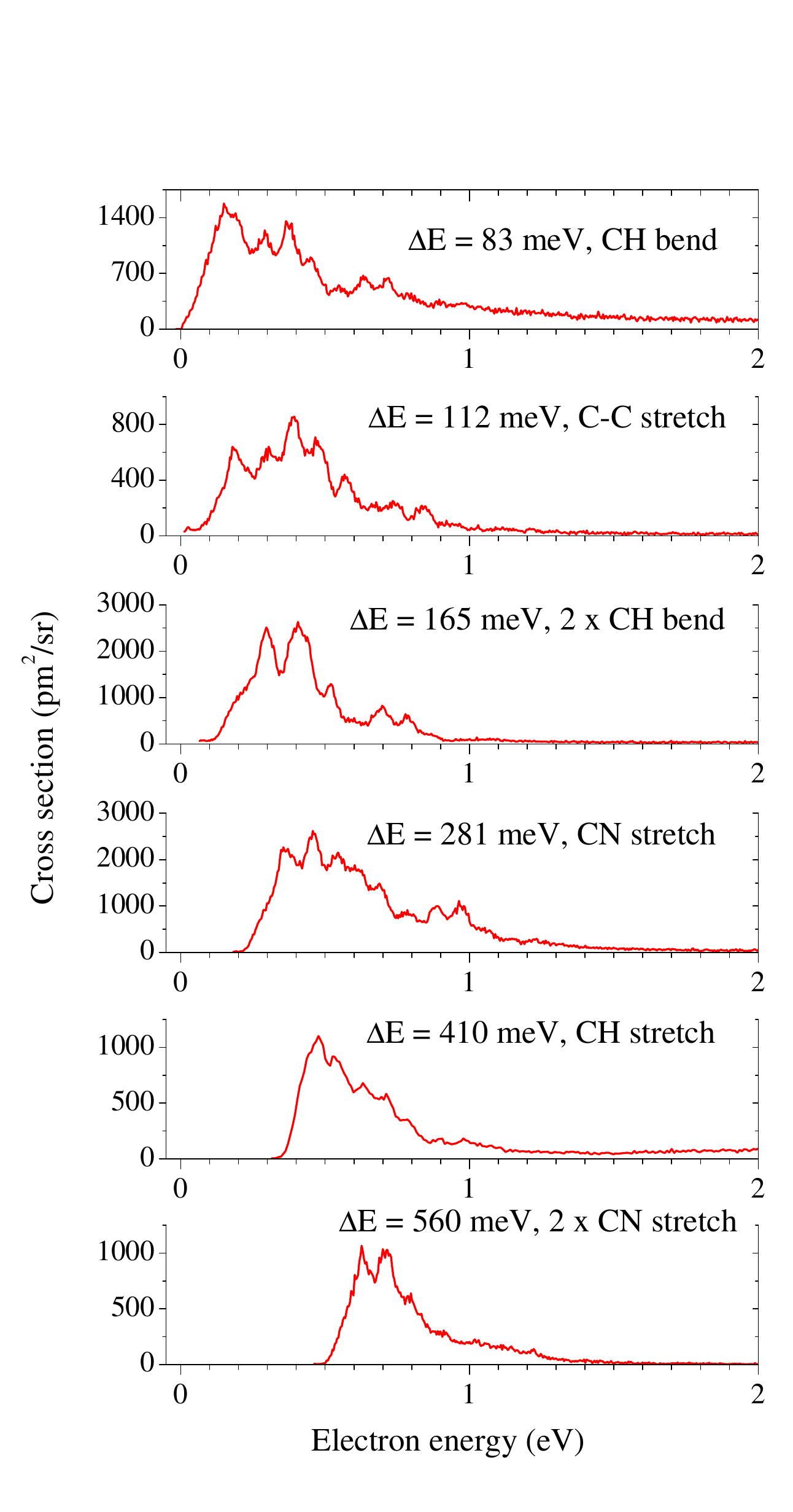}
\caption{The vibrational excitation cross sections for individual vibrations in HC$_3$N as functions of incident electron energy with the low-energy horizontal scale expanded.}
\label{fig:eds_short}
\end{figure}

\begin{figure}[tb]
\includegraphics[width = 7cm]{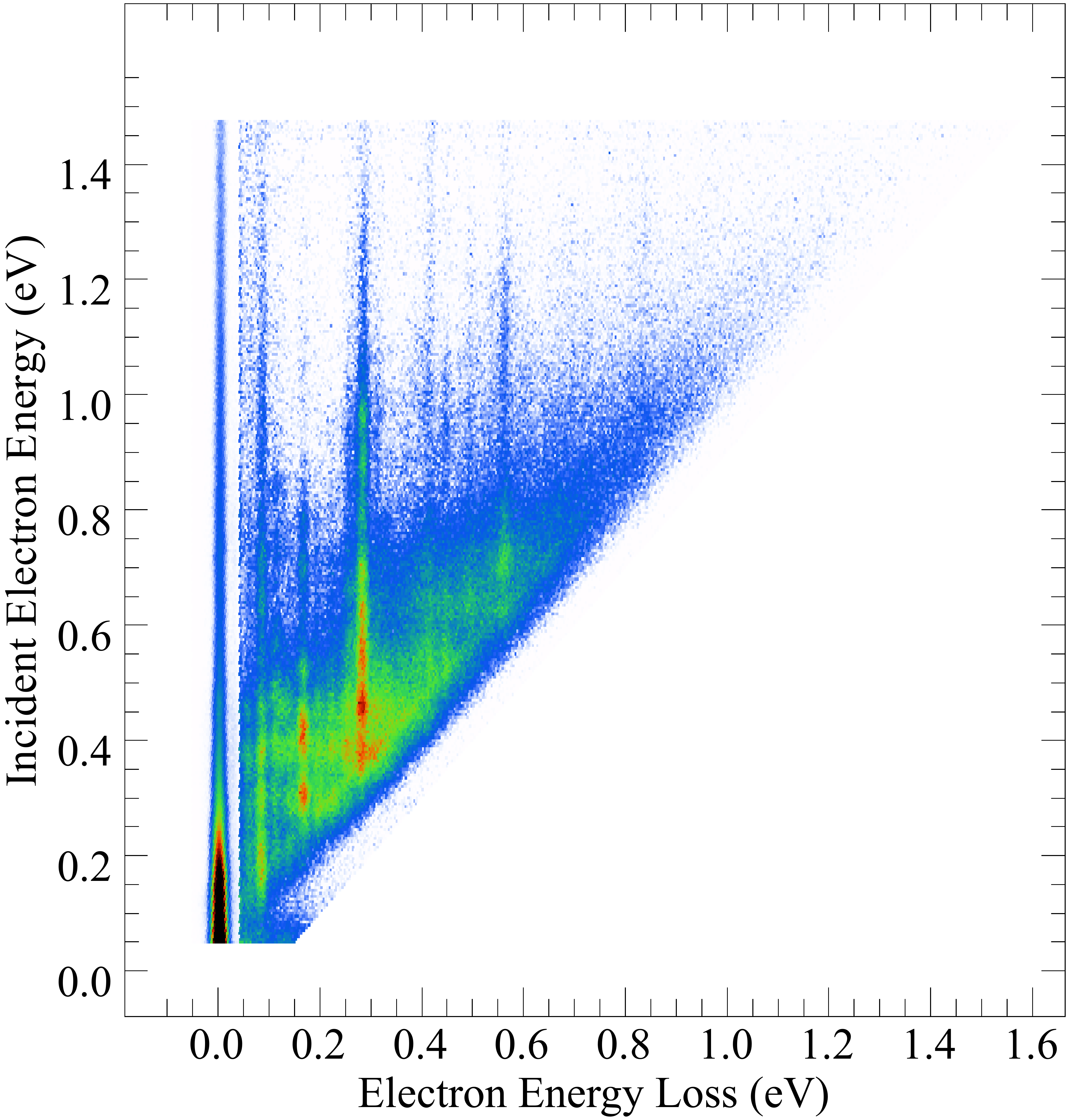}
\caption{Two-dimensional electron energy loss spectrum of HC$_3$N. The intensity of the elastic peak (energy loss = 0 eV) is reduced by a factor of 20 with respect to the rest of the spectrum.}
\label{fig:2d}
\end{figure}

\begin{figure}[tb]
\includegraphics[width = 7cm]{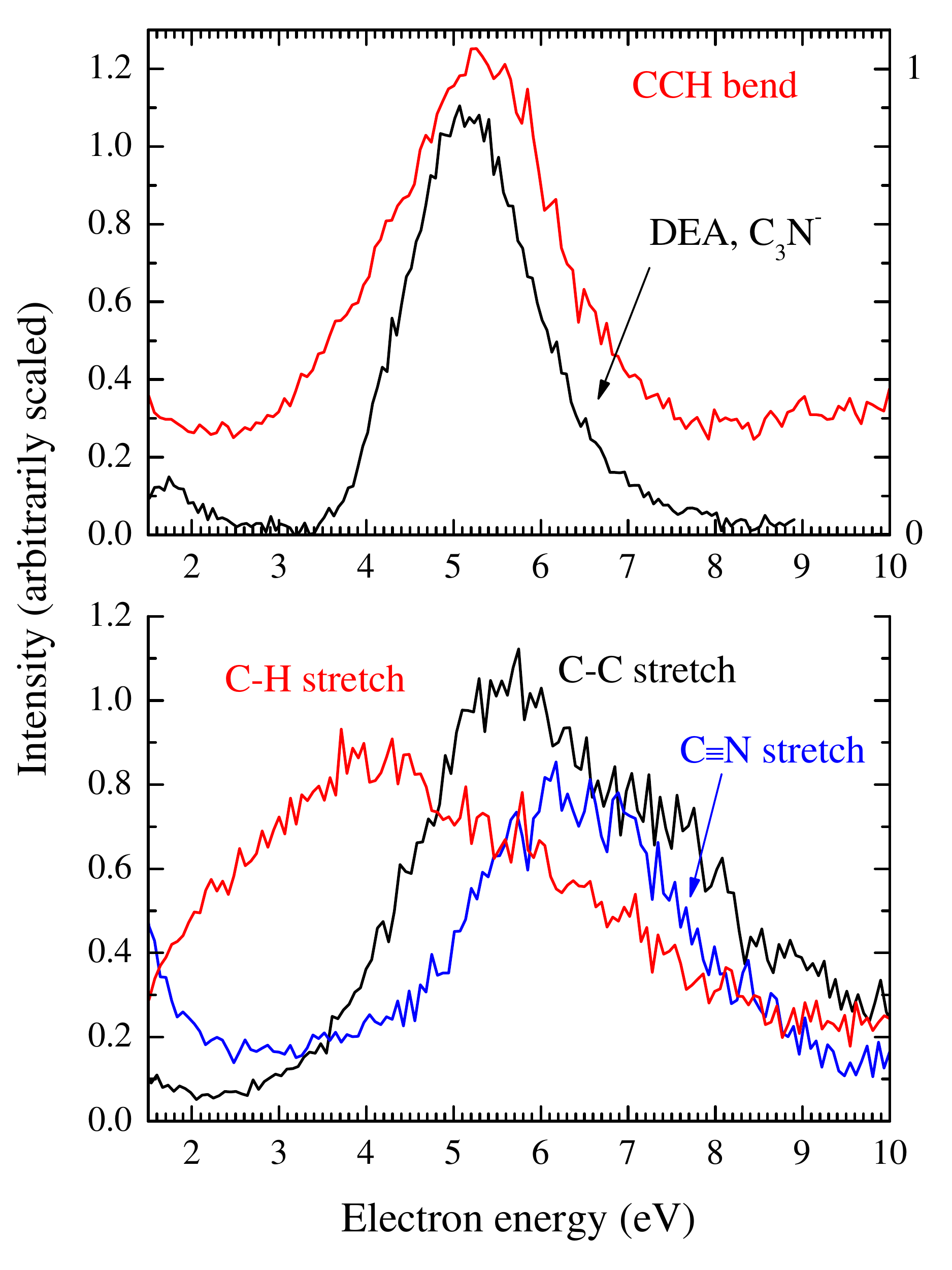}
\caption{High-energy part of the individual vibrational excitation excitation cross sections and DEA C$_3$N$^-$ ion yield. The raw data from figure~\ref{fig:eds_long} are multiple times reduced (neighbouring channels are averaged). For the sake of this comparison, the data are arbitrarily scaled.}
\label{fig:eds_comp}
\end{figure}

Figure~\ref{fig:eds_long} shows the excitation curves of the individual vibrations. Here, the energy difference between the monochromator and analyzer is kept constant and both are being scanned. Such excitation curves are a sensitive probe for the formation of resonances: if a temporary anion is formed at certain incident electron energy, the probability of energy transfer to nuclear motion (= vibrational excitation) strongly increases. %The vibrational modes that are primarily excited reflect the dissociative direction of the resonant potential energy surface at the point of the electron attachment.  
The observed bands can be divided into two groups, the narrow ones at low-energies, approximately below 1 eV and much broader bands at higher energies, above 2~eV. The low-energy part of the spectra is separately shown in figure~\ref{fig:eds_short} and in the form of a two-dimensional spectrum in figure~\ref{fig:2d}. The high-energy part (with the reduced number of channels) is shown rescaled in figure~\ref{fig:eds_comp}.

Let us first focus on the high-energy part. The dominant contribution to the excitation of all vibrations seems to originate from the formation of $\pi_2^*$ resonance, however, clear differences in the excitation of individual modes are demonstrated in figure~\ref{fig:eds_comp}. Since the two $\sigma^*$ resonances are dissociative along the molecular axis and will probably excite the bending vibrations only negligibly, we presume that the ``true'' shape of the $\pi_2^*$ resonance is demonstrated by the CCH bend excitation curve (top panel of figure~\ref{fig:eds_comp}). This places the center of the $\pi_2^*$ resonance to 5.3~eV. 

The C-H stretch vibration has the maximum clearly shifted to lower energies. The $\sigma_1^*$ orbital (table~\ref{tab:orbitals}) has the largest coefficient on the corresponding carbon and hydrogen atoms and an antibonding character along this bond. We conclude, that the C-H stretch vibration is the only one, which is influenced by the formation of the broad $\sigma_1^*$ resonance with the center around 4~eV. 

The C$\equiv$N vibration excitation curve is shifted to higher energies when compared to the CCH bend. This is caused by the formation of the $\sigma_2^*$ resonance with a strong antibonding character across the C$\equiv$N bond. This resonance is also visible in the excitation of the C-C stretch mode as the right shoulder superimposed on the dominant $\pi_2^*$ resonance.

We now turn to the low energy part of the vibrational excitation spectra shown in detail in figures~\ref{fig:eds_short} and~\ref{fig:2d}. The excitation curves have peculiar shapes. This is caused by an interplay of two effects. The first one is related to the strong dipole moment of cyanoacetylene (3.72~Debye) which is expected to lead to threshold peaks in the vibrational excitation cross sections. Such peaks, first observed in hydrogen halides~\cite{rohr76} are common in all polar molecules. The second effect is the formation of the $\pi_1^*$ resonance around 0.5~eV. The small width of the resonance leads to a pronounced boomerang structure, visible in all vibrational modes. %This type of structure has been interpreted for the first time by Herzenberg to 
The boomerang structure originates from the vibrational motion of the nuclear wavepacket on the anion potential energy surface. Due to the long lifetime of the resonant state, the nuclei will undergo several vibrations prior to the electron detachment. The oscillatory structure originates from the interference of the outgoing and returning nuclear wavepacket.~\cite{herzenberg71} It is commonly manifested  as a structure on top of a vibrational excitation band. The present accidental overlap of the $\pi^*$ resonance and the threshold peak causes the rather exotic accumulation of the boomerang structure on the falling edge of the peak.

The present data enable to judge the accuracy of different methods used to calculate the resonant energies in table~\ref{tab:orbitals}. So far, the only experimental data on these states came from the DEA spectroscopy~\cite{graupner06}. Those are, however, influenced by energetical threshold cutoffs, or by the formation of core-excited resonances. The present data enable an unambiguous determination of the position of the $\pi_2^*$ resonance at 5.3~eV. This compares surprisingly favorably with the value obtained from the scaling formula (5.5~eV) and reasonably well with the CAP value of 6.2~eV. The single-center expansion scattering calculation~\cite{sebastianelli12} overestimates the position of this resonance by almost 3~eV (8.18~eV). For the $\pi_1^*$ resonance, the determination of the experimental center is complicated due to overlap with the threshold peak, however, judging from the boomerang structure in the C-C stretch and C-H bend excitations (figure~\ref{fig:eds_short}), the center can be placed to 0.5~eV. Again, the CAP method predicts this resonance better than the two scattering calculations (0.7~eV vs. 1.94 and 1.51~eV). These two also overestimate the energy of the $\sigma_2^*$ resonance, which has the experimental center between 6 and 7~eV, judging from the C-N stretch excitation curve in figure~\ref{fig:eds_comp}.

A further insight into the low-energy part can be gained from the two-dimensional spectrum in figure~\ref{fig:2d}. 2D electron energy loss spectrum~\cite{regeta13} is a collection of many energy loss spectra recorded at various incident energies. It provides a complete picture of the vibrational nuclear dynamics. A horizontal cut through such spectrum corresponds to an energy loss spectrum, such as shown in figure~\ref{fig:eels}, a vertical cut corresponds to an excitation curve of a given energy loss, such as shown in figure~\ref{fig:eds_short}. The diagonal line $E_i = \Delta E$ is the threshold line, corresponding to the outgoing electrons with zero kinetic energies. 

The 2D spectrum agrees fully with the individual vibrational cross sections. Additionally, it reveals one more feature: approximately above 0.2~eV incident electron energy, the electrons along the diagonal ($\Delta E = E_i$) form a weak continuous stripe instead of appearing only at the sharp energies of individual vibrations. These electrons are ejected with residual energies close to zero, independent of their incident energy. Note, that the analyzer has a low transmission of electrons with residual energies below some 30~meV to 50~meV, hence the threshold signal appears somewhat higher than  $E_r = 0$~eV. It is also visible as the high background signal in the energy loss spectrum on the upper panel of figure~\ref{fig:eels}.%In the energy loss spectra in figure~

These threshold electrons can be interpreted using the potential energy surfaces of Sommerfeld and Knecht.~\cite{sommerfeld05} According to their calculations, cyanoacetylene posseses a valence-bound anion, however, it's equilibrium geometry is far from the neutral one.  It has a trans-bent zig-zag structure, however, with an adiabatic electron affinity close to zero. Apart from this, HC$_3$N supports a dipole-bound state %with the equilibrium geometry of the neutral molecule 
with the potential energy curve lying several meV below the neutral one, it's equilibrium geometry thus corresponds to the neutral's linear structure.  
The linear transit between the two anion states (valence and dipole-bound) shows a barrier of approximately 0.2~eV. The origin of the slow electrons is thus following: if an electron with the incident energy $E_i > 0.2$~eV is captured in the low-lying $\pi^*$ resonance, the nuclear framework starts to move towards the geometry of the valence-bound anion, distorting the linear structure towards the trans-cis bent one. As soon as the geometry gets to the point where the anion surface lies below that of the neutral, the electron detachment is suppressed: it is energetically impossible for the electron to detach. However, the excess energy is stored in the nuclear degrees of freedom and efficiently randomizes over the vibrational degrees of freedom. The motion on the electronically bound part of the potential surface is statistical, so the nuclei may again get to the configuration, where the valence anion energy lies above that of the neutral. %Here the excess energy 'thermalizes'
At this crossing point of the neutral and the anion surface the electron is unbound again and can detach. A number of previous examples~\cite{allan_habil89, allan_formic_prl07, allan_feco18} shows that such electrons detach basically as soon as they can and are thus emitted with close-to-zero residual energies. %Incomplete randomization...

%\subsection{2-Dimensional spectrum}
\begin{figure}[tb]
\includegraphics[width = 7cm]{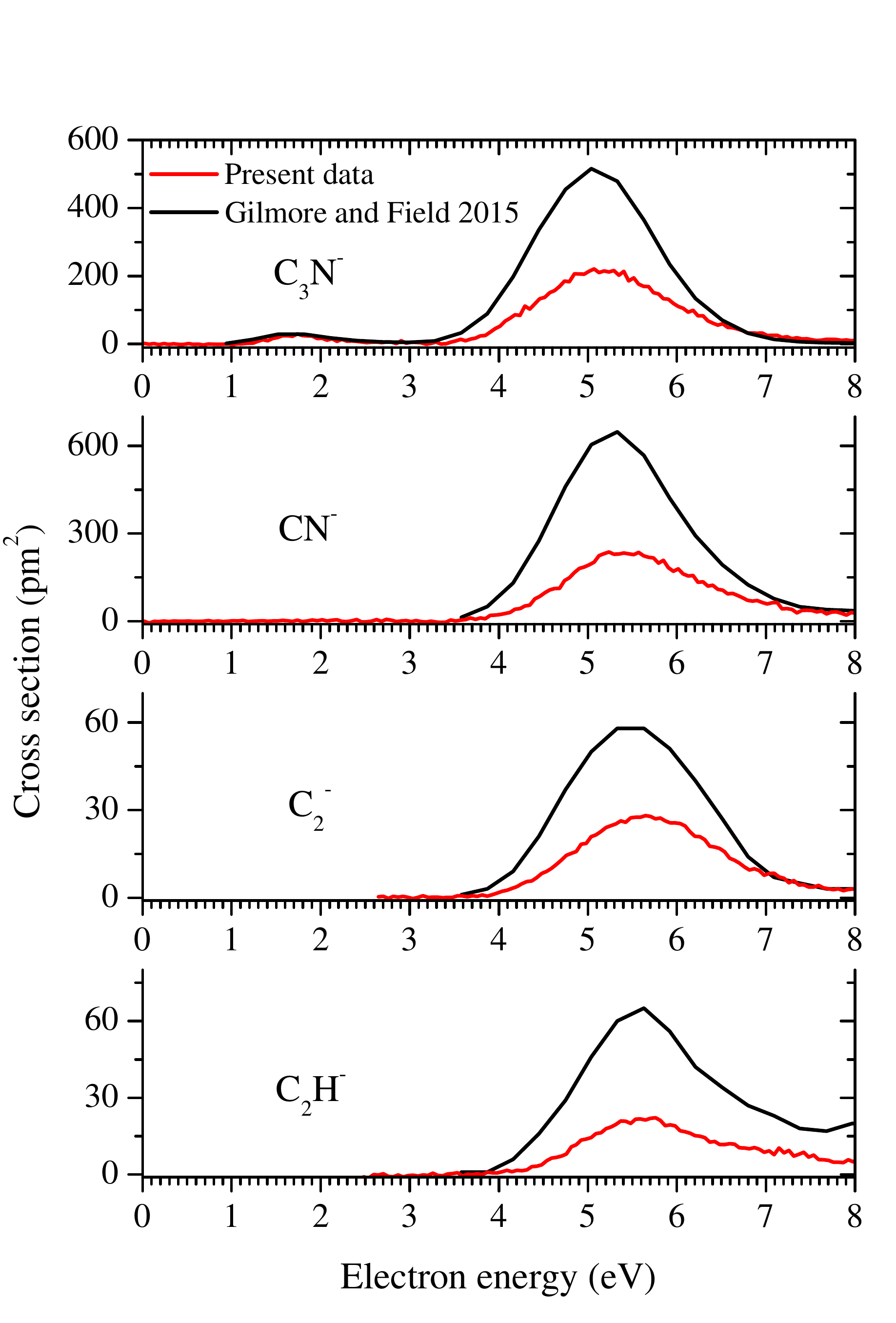}
\caption{Partial DEA cross sections for the production of all anionic fragments from HC$_3$N. Red lines: present data, black lines: Gilmore and Field.~\cite{gilmore15} }
\label{fig:dea_all}
\end{figure}

\begin{figure}[tb]
\includegraphics[width = 7cm]{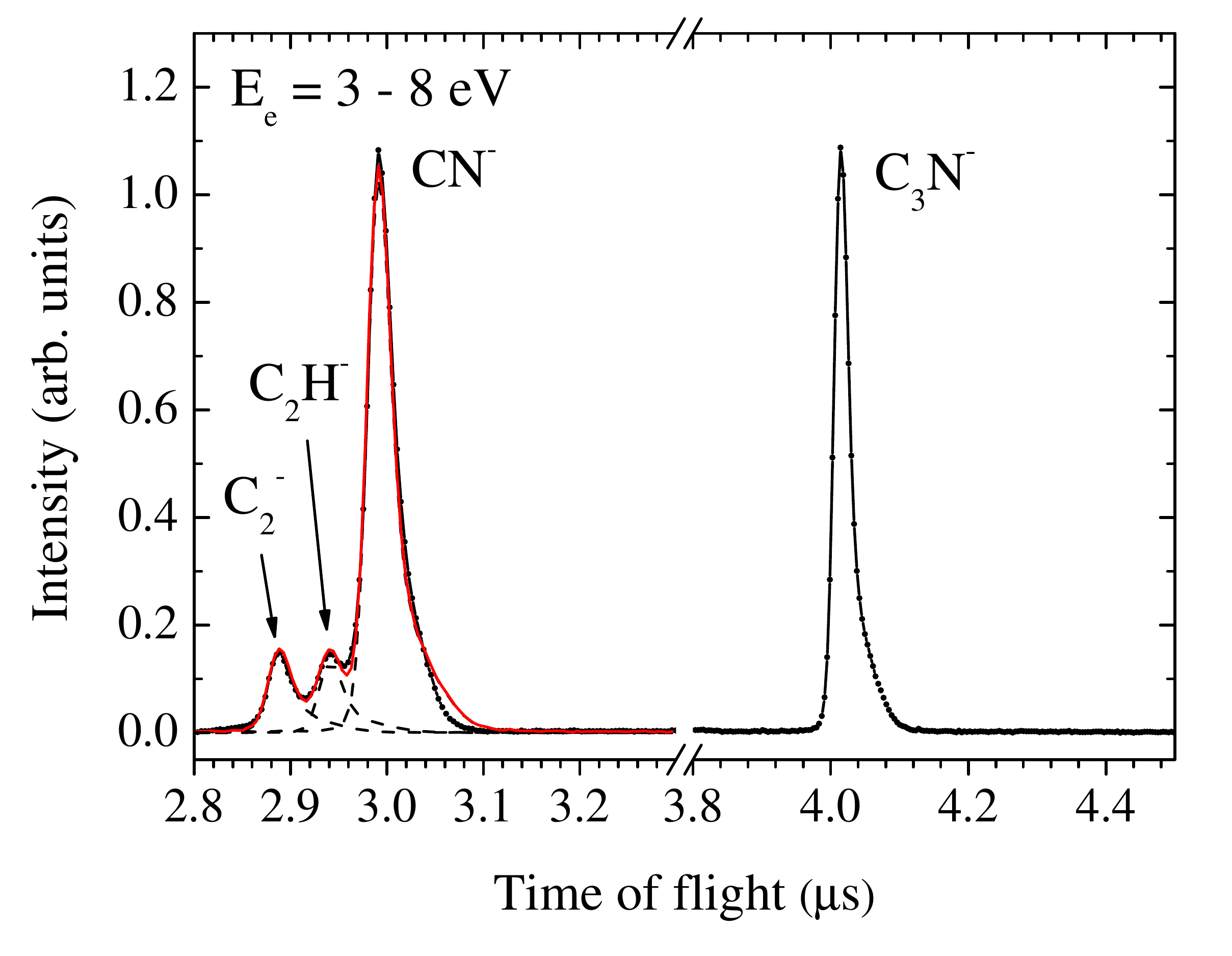}
\caption{Cumulative negative ion time-of-flight spectrum in the energy range 3 to 8~eV. Lines with points: experimental data, dashed lines: fitted contributions from the peaks with mass to charge ratios 24, 25 and 26, red line: sum of the individual contributions.}
\label{fig:MS}
\end{figure}

\begin{figure}[tb]
\includegraphics[width = 7cm]{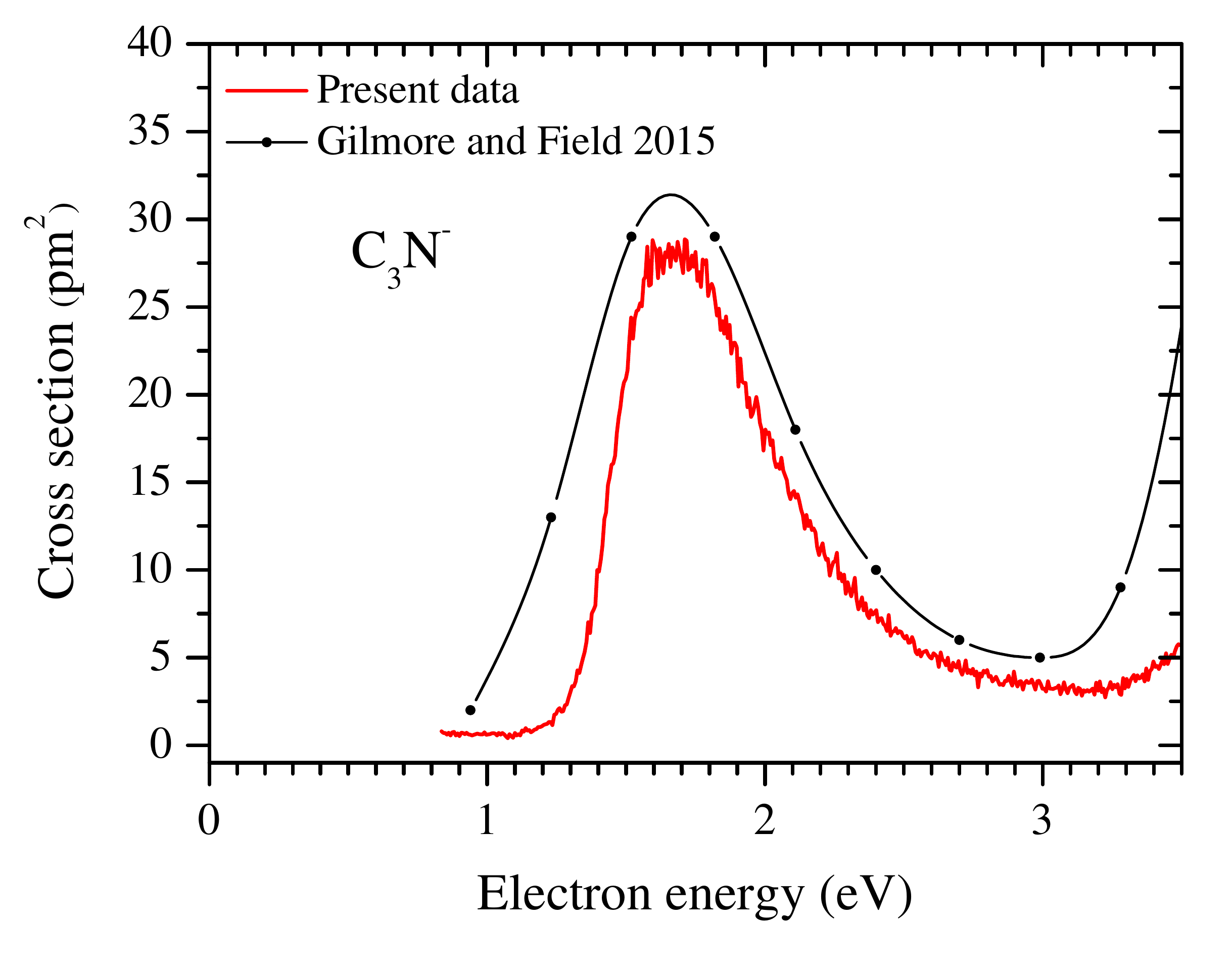}
\caption{Low-energy DEA band for the C$_3$N$^-$ fragment. Red line: present data, black points: Gilmore and Field.~\cite{gilmore15} }
\label{fig:dea_m50}
\end{figure}

\subsection{Dissociative electron attachment}
Figure~\ref{fig:dea_all} shows the absolute cross section for the production of individual fragment anions from HC$_3$N. The recent data of Gilmore and Field~\cite{gilmore15} are shown for comparison. The two data sets show an excellent agreement concerning the shapes of the individual DEA bands. However, there is a consistent quantitative disagreement. % in absolute values and certain variation of the branching ratios. 
%b: the present cross sections are systematically by a factor of XXX smaller. This may be somewhat less apparent for the low-lying DEA band in C$_3$N$^-$ production, which seems to have a similar height than the reference data. However, as demonstrated in the figure~\ref{fig:dea_m50}, where this band is shown separately, the present experiment has a better energy resolution. The ratio of the energy-integrated cross sections (invariant of the beam resolution) is XXX, consistent with other fragments.
We will use the energy-integrated cross section $\sigma_I$ (invariant of the beam resolution) for the discussion. 
On the main DEA band, spanning between 3 and 8~eV, the ratio of our $\sigma_I$ for the C$_3$N$^-$ production (411~eV pm$^2$) to that of Gilmore and Field is 0.47.
This disagreement is more or less consistent for all the four fragments.%, except for CN$^-$. 
The present branching ratio between the fragments C$_3$N$^-$:C$_2^-$:C$_2$H$^-$:CN$^-$ are 1:0.14:0.12:0.95. The branching ratio of Gilmore and Field are 1:0.13:0.15:1.33, they thus agree very well, apart from the CN$^-$ which had higher abundance in the measurements of Ref.~\onlinecite{gilmore15}. At this point, it should be noted that our time-of-flight analyzer does not fully resolve the three fragments with mass-to-charge ratios 24, 25 and 26. When designed,~\cite{may_acet09} the resolution has been compromised to the fact that the setup is quantitative. There are for example no grids separating the two acceleration regions. This on one hand distorts the Wiley-McLaren type time focusing, on the other hand it means undisturbed transmission of extracted anions. Still, as is illustrated in figure~\ref{fig:MS}, the mass resolution is high enough to determine the branching ratios between the three fragments reliably. The spectrum is cumulative~\cite{lengyel_beilstein17, lengyel_hno3_17} - it has been obtained as a sum of the mass spectra in the energy range 3 to 8~eV. The dashed lines show the individual contributions of the three close-lying fragments %in the energy range 3 to 8~eV 
and the full red line shows their sum.   

Somewhat surprisingly, the quantitative level of agreement between the present data and those of Gilmore and Field is better for the first DEA band in the C$_3$N$^-$ production, shown magnified in figure~\ref{fig:dea_m50}. The ratio of the energy-integrated cross sections of this band is 0.68. 

The probable origin of the quantitative discrepancy are the different calibration methods used to obtain the absolute values. Gilmore and Field used the O$^-$ signal from the background water vapor for the cross calibration. The ratio of the HC$_3$N/H$_2$O number densities was obtained from the recorded ion yields of the positive ions and their calculated absolute cross sections in the BEB formalism. Considering this rather indirect approach, the present agreement of the absolute cross section within a factor of two can be actually viewed as very good. Both experiments have quoted uncertainty of $\pm$ 20\% and the difference between the absolute values is only slightly larger than the combined error limits. Due to more direct calibration procedure, the present values might be considered to be more reliable.

The comparison with the vibrational excitation cross sections brings new light on the DEA mechanism. As seen in figure~\ref{fig:eds_comp}, the band at 5.5~eV is very similar the shape of CCH bend excitation cross section, which suggests that the DEA is mediated by the formation of the $\pi_2^*$ resonance. 
Graupner et al.~\cite{graupner06} did the same assignment, however, since their reference center of the $\pi_2^*$ resonance was that calculated by Sommerfeld and Knecht~\cite{sommerfeld05} at 6.2~eV, they had to argue with a survival probability shift in order to explain the different DEA peak position. The current comparison in figure~\ref{fig:eds_comp} shows that the DEA band actually overlaps with the $\pi_2^*$ resonance very well. 

Still, there is one aspect which invokes caution with this assignment, and this is the large width of the $\pi_2^*$ resonance. The corresponding bands (both in DEA and in vibrational excitation spectra) are approximately 2~eV broad. The width of the band is determined by two factors: (i) the autodetachment width $\Gamma$ and (ii) the projection of the nuclear wavefuction on the resonant state (reflection principle). Anyhow, such broad bands suggest, that $\Gamma$ itself is rather large, in agreement with the theoretical calculations which evaluated it to be 1.1~eV (Ref.~\onlinecite{sommerfeld05}) or 0.76~eV (Ref.~\onlinecite{sebastianelli12}). From the uncertainty principle, a resonance width of 1~eV corresponds to the lifetime towards electron autodetachment of 0.3 femtoseconds. It is somewhat surprising that such a short-lived state gives rise to rather high dissociative cross section. An alternative origin of the DEA yield would be a core-excited resonance: neutral HC$_3$N posses electronically excited states ($^1\Delta_u$) lying between 5.5 and 6.2~eV.~\cite{ferradaz09} Assuming a typical stabilization energy of 0.4~eV, the corresponding Feshbach resonance would be located exactly around the present DEA band. Such resonances are typically very narrow and are not visible in the vibrational excitation cross section.~\cite{allan_habil89} They also typically lead to rich fragmentation pattern.~\cite{janeckova_thf14, zawadzki_pyruvic18} The agreement in figure~\ref{fig:eds_comp} thus might be coincidental.
It is worth noting that a similar dispute, whether the dominant DEA band is caused by an accidentally overlapping shape $\pi^*$ or a core-excited resonance, has appeared for diacetylene C$_4$H$_2$.~\cite{allan_diac11, curik_diac14} 

Only the C$_3$N$^-$ fragment, created by the hydrogen abstraction, is observed at lower energies with the peak at 1.7~eV. It was shown by calculating the threshold energies~\cite{graupner06} that other channels are energetically closed in this energy range. The threshold for the C$_3$N$^-$ production is 1.37 $\pm$ 0.2~eV %skontroluj neumarka
which is causing a sharp onset of the present cross section in figure~\ref{fig:dea_m50}. Two effects can in principle contribute to the origin of this band.

(i) As assigned previously~\cite{graupner06}, it can originate from a high-energy shoulder of the $\pi_1^*$ resonance (the center of the resonance lies considerably below the threshold energy). This seemed very reasonable, since this resonance is rather narrow so it would lead to high survival factor. However, as can be seen in figure~\ref{fig:eds_short}, all the cross sections for the vibrational excitation are diminishing above 1.3~eV, so the DEA band seems to have almost no overlap with the $\pi_1^*$ resonance. 

(ii) The second option is that the DEA proceed via formation of the $\sigma^*$ resonance, whose lower tail overlaps with the DEA band as can be seen in the C-H stretch vibrational excitation in figure~\ref{fig:eds_long}. Judging from a large width of such resonance alone, it should lead to negligible DEA cross section, since all electrons would autodetach. However, it is now well established, that in molecules with large dipole moments (or even nonpolar molecules with high polarizabilities), the dissociative cross section of $\sigma^*$ resonances can reach very high values. The interaction of dipole bound (or virtual states) with the pure $\sigma^*$ states suppresses the autodetachment channel.  
The cyanoacetylene's  dipole moment of 3.72 Debye opens this possibility. It should be however noted, that such dipole-supported $\sigma^*$ resonances often lead to sharp structures in the DEA cross section. These structures - downward steps or even  oscillations - appear at the opening of the new vibrational excitation channels in the direction of the dissociating bond, in this case the C-H vibration. Taking into account the anharmonic vibrational levels,~\cite{mallinson76} the 0$\to$4 transition in C-H stretch vibration is open at 1.56~eV and 0$\to$5 transition at 1.94 eV. No such structures are visible in figure~\ref{fig:dea_m50}. It should be noted that the DEA spectra of molecules like hydrogen halides~\cite{fedor_hbr07, fedor_hbr08, fedor_hcl10} or formic acid~\cite{janeckova_formic13} do show discernible structures at electron beam resolution comparable to the present one (approximately 100 meV). 

There seems to be no unambiguous evidence for any of the two mechanisms to be prevalent in the dehydrogenation DEA around 1.7~eV. Our recent results for the HNCO molecule~\cite{zawadzki_prl18} even suggest that often there is even no sharp distinction between these possible: upon any out-of-line geometry distortion the $\pi^*$ and $\sigma^*$ states mix and the actual dissociation mechanism is given by the interplay of them.

\section{Conclusions}
In conclusion, we have probed the resonances in cyanoacetylene by measuring cross sections for elastic electron scattering, vibrational excitation and dissociative electron attachment. Data from these three scattering channels are mutually consistent and provide information about both the non-dissociative and dissociative nuclear dynamics on the transient anion potential surfaces. 

Several effects influence the probed electron-induced processes. %Several effects are manifested in the results.
 One is the strong dipole moment of HC$_3$N which is manifested as the low-energy peak in the elastic scattering and as the threshold peaks in all the vibrational excitation channels. The second dominating effect is the formation of four resonances. The lower $\pi_1*$ resonance is the narrowest and its long lifetime leads to pronounced boomerang oscillatory structures in the vibrational excitation cross sections. At higher electron energies, the formation of the broad $\sigma_1^*$ and $\sigma_2^*$ resonances is reflected in the vibrational excitation cross sections of the C-H stretch and C$\equiv$N stretch modes, while the CCH bending mode excitation is probably exclusively mediated by the formation of the $\pi_2^*$ resonance. This resonance also dominates the DEA spectrum and it leads to production of four anionic fragments. 
The existence of the bound HC$_3$N$^-$ anion and the crossing of its potential energy curve with that of the neutral molecule (boundary between the resonant and the bound state) is manifested by the threshold signal in the two dimensional energy loss spectrum. Here the electrons are emitted with close-to-zero residual energies independent of the incident energy, which is caused by the randomization of the vibrational motion on the bound anion surface.

\section*{Acknowledgments}
This work is part of the project Nr. 17-04844S of the Czech Science Foundation. L. B. acknowledges support from the FWF project DK-ALM:W1259-N27, M. P. and J. \v{Z}. acknowledge partial support from CSF project Nr. 17-14200S. We wish to thank Roman \v{C}ur\'{i}k, Prague, for numerous discussions of resonances and of this manuscript. 

\bibliographystyle{apsrev}
\bibliography{HC3N}
%\begin{thebibliography}{99}

%\end{thebibliography}

\end{document}